# Development of very high $J_c$ in Ba(Fe$_{1-x}$Co$_x$)$_2$As$_2$ thin films grown on CaF$_2$

C. Tarantini[1*], F. Kametani[1], S. Lee[2†], J. Jiang[1], J.D. Weiss[1], J. Jaroszynski[1], E.E. Hellstrom[1], C. B. Eom[2], D.C. Larbalestier[1]

[1] Applied Superconductivity Center, National High Magnetic Field Laboratory, Florida State University, Tallahassee FL 32310, USA

[2] Department of Materials Science and Engineering, University of Wisconsin-Madison, Madison, WI 53706, USA

Ba(Fe$_{1-x}$Co$_x$)$_2$As$_2$ is the most tunable of the Fe-based superconductors (FBS) in terms of acceptance of high densities of self-assembled and artificially introduced pinning centres which are effective in significantly increasing the critical current density, $J_c$. Moreover, FBS are very sensitive to strain, which induces an important enhancement in critical temperature, $T_c$, of the material. In this paper we demonstrate that strain induced by the substrate can further improve $J_c$ of both single and multilayer films by more than that expected simply due to the increase in $T_c$. The multilayer deposition of Ba(Fe$_{1-x}$Co$_x$)$_2$As$_2$ on CaF$_2$ increases the pinning force density ($F_p=J_c\times\mu_0 H$) by more than 60% compared to a single layer film, reaching a maximum of 84 GN/m$^3$ at 22.5T and 4.2 K, the highest value ever reported in any 122 phase.

[*] e-mail: tarantini@asc.magnet.fsu.edu

[†] Current address: School of Materials Science and Engineering, Gwangju Institute of Science and Technology, Gwangju 500-712, Republic of Korea

In the last 6 years the Fe-based superconductors (FBS) have been widely studied because of their many existing compounds, their high critical temperature $T_c$ up to 55K,[1,2,3,4] large upper critical field $H_{c2}$ exceeding 100 T and relatively low anisotropy.[5,6,7,8,9] Interest in these materials has also been amplified by both their unconventional coupling mechanisms[10,11,12] and their potential for high field applications.[13,14,15] In fact, besides having a high $H_{c2}$, FBS have also shown high intragrain critical current density $J_c$.[16,17] Despite the relatively low $T_c$ (about 22 K in the bulk form), Co-doped BaFe$_2$As$_2$ (Ba122) has been studied by several groups[18,19,20,21,22] because thin film growth is relatively easy compared to other FBS materials. We previously demonstrated that Ba122 thin films have the impressive capability to accept a high density of secondary phases that act as strong pinning centres.[23,25] Self-assembled BaFeO$_2$ (BFO) nanorods parallel to the *c*-axis enhance $J_c$ in this crystallographic direction reducing the critical current anisotropy,[23,24] whereas artificially introduced defects, obtained by multilayer deposition, improve $J_c$ not only along the *ab*-planes but in the entire angular range.[25] The strong effectiveness of these defects is due to their typical sizes comparable to $2\xi_0 \sim 5$ nm, where $\xi_0$ is the superconducting coherence length at 0 K. The volume fraction of precipitates, nanoparticle and/or nanorods in Ba122 can exceed 15 vol.%, much higher than in the YBa$_2$Cu$_3$O$_{7-x}$ (YBCO) case where more than 5 vol.% produces a suppression of the matrix superconducting properties.[26,27,28,29] In Co-doped Ba122 (Co-Ba122), the highest pinning force density $F_p$ obtained so far has been about 53 GN/m$^3$ at 4.2 K,[25] whereas $F_p$ reaches 59 GN/m$^3$ in P-doped Ba122,[30] whose $T_c$ in bulk form is ~30-31 K.[31,32]

In the last few years a general improvement in thin film performance has been obtained by growing FeSe$_{1-x}$Te$_x$,[33] SmFeAs(O,F),[34,35] NdFeAs(O,F),[36] and Co-Ba122 [37] on CaF$_2$ substrates. This improvement is related to a compressive strain along the *a*-axis and its origin has been attributed to two possible effects. In ref. 37 the strain was correlated to a substrate/film lattice mismatch that, in the case of Co- Ba122, reaches 2.7%: in this work $T_{c,0}$ ($T_c$ at zero-resistance) was enhanced to more than 24 K and the self-field $J_c$ to ~1MA/cm$^2$ at 10 K. Differently in ref. 38 the origin of the film strain was recently ascribed not to the direct lattice mismatch but to a more indirect thermal expansion mismatch due the larger thermal expansion coefficient of $CaF_2$ with respect to Co-Ba122 and other substrates. In order to study the effect of $CaF_2$ substrate on samples with modified pinning properties, we investigated single- and multi-layered Co-Ba122 thin films grown on $CaF_2$ and on $SrTiO_3$-templated $(La,Sr)(Al,Ta)O_3$ (LSAT/STO).[39] We characterized these films over a wide temperature range and in high magnetic fields up to 45T at the National High Magnetic Field Laboratory. We found that the critical temperature of the films on $CaF_2$ is enhanced by about 3 K with respect to similarly grown films on LSAT. In the multilayer case, $J_c$ and $F_p$ increase above what is expected for the simple $T_c$ enhancement, reaching a $F_p$ maximum of 84 $GN/m^3$ at 22.5T and 4.2 K, 60% larger than the corresponding single layer film and 2-3 times more effective than multilayer deposition on LSAT/STO. In order to understand the origin of this enhancement, we also performed transmission electron microscopy (TEM) and electron energy loss spectroscopy (EELS) characterizations that clearly show different microstructures than in previous films. We discuss how this strong enhancement is related to the global change in the microstructure and to the different nature of pinning centres with an important contribution by point defects in films on $CaF_2$.

## Results

**Samples and microscopic characterization**

Two different types of epitaxial thin films were deposited by pulsed laser deposition on (001) $CaF_2$: one was a Co-Ba122 single layer film (named CaF-S), one a multilayer film (CaF-M) prepared by alternating Co-Ba122 and BaO-doped Co-Ba122 layers (see Methods for details). These two samples were compared with previously characterized single and multilayer films deposited on LSAT/STO: the single layer (LSAT-S) was grown similarly to CaF-S, whereas the multilayer (LSAT-M) was grown by alternating Co-Ba122 and undoped Ba122 layers (see Methods). $T_{c,0}$ values are 22.1-22.8 K for the films on LSAT and 25.4-26.0 K for the films on $CaF_2$ which is an increase of more than 3 K, confirming the positive strain effect of the substrate. The main properties of the samples are summarized in Table I. All samples have good and comparable crystalline structure as determined by x-ray diffraction with small out-of-plane and in-plane full width at half maximum (FWHM): we determined $\Delta\omega < 0.5°$ and $\Delta\phi \sim 0.6°$ for the samples on $CaF_2$ and $\Delta\omega < 0.4°$ and $\Delta\phi \sim 0.7°$ for the samples on LSAT.[39]

The cross-sectional TEM image of CaF-S (Fig. 1a) reveals that the single layer film on $CaF_2$ has a remarkably clean Co-Ba122 layer: only threading dislocations that extend through the whole thickness and a thin transition layer at the interface between the Co-Ba122 layer and substrate are observed. No nanorods or particles of secondary phases were found in the Co-Ba122 layer. The estimated density of these dislocations corresponds to a matching field $B_\phi(H//c)$ of less than 3 T. On the other hand, as shown in Fig. 1b and c, the multilayer CaF-M has short *c*-axis nanorods, *ab*-aligned precipitates and round nanoparticles similar in sizes and shapes to the ones we have found in the LSAT films (LSAT-M).[25] However the density of these nanorods and precipitates is smaller than in LSAT-M and their distribution is clearly not uniform through the thickness. In fact a notable decrease of the defect density is seen near the substrate (Fig. 1b) with a matching field $B_\phi(H//ab)$ that ranges from ~3.2 T near the substrate to ~5.7 T close to the top surface (in the LSAT-M film, $B_\phi$ is 8-9 T). An EELS linescan across the precipitates suggested that they are Ba-O. Another feature of the CaF-M film is a thicker reaction layer at the Co-Ba122/substrate interface compared with the CaF-S film.

In order to map the chemical composition, energy filtered TEM (EFTEM) using EELS was performed on the CaF-M film (Figure 2). The Ca map of Fig. 2a confirms that the Co-Ba122 layer is Ca-free, except for a transition layer. The Fe and As distributions (Fig. 2b and d) are uniform from the substrate interface (represented as the white dotted line) to the top of the Co-Ba122 layer. As is seen in Fig. 2e, Ba is also uniform in the Co-Ba122 layer, but there is a clear depletion at the transition layer (just above the interface). Fig. 2c reveals the presence of oxygen in the transition layer and the substrate especially near the interface; moreover bright spots of oxygen in the Co-Ba122 layer correspond to the location of precipitates, confirming that they contain oxygen. Surprisingly, no fluorine was detected either in the film or in the top part of the substrate (the absence of fluorine in the substrate was verified down to at least 500nm below the interface). The elemental mapping suggests that the transition layer is mainly

Fe-As-O with some slight interdiffusion of Ba and Ca, and that the top part of the $CaF_2$ substrate actually transformed into calcium oxide during deposition, as judged by the absence of fluorine and clear segregation of oxygen on both sides of the interface.

**Transport characterization**

The critical current density $J_c$ was measured up to 16 T over a wide temperature range (see Methods) with magnetic field applied along the two main crystallographic directions (H//$c$ and H//$ab$). Figure 3 (left panels) compares the results of all four films at 16K. Both films on $CaF_2$ exceed 1 MA/cm$^2$ at 16 K and self-field, more than twice the $J_c$(16K,s.f.) of the films on LSAT. The in-field properties are also significantly improved: for instance, the irreversibility field $\mu_0 H_{Irr}$ (determined at $J_c$=100A/cm$^2$) with H//$c$ increases from 10-11 T in the LSAT-samples to 15.5-17.5 T in the CaF-samples. With H//$ab$, $\mu_0 H_{Irr}$ largely exceeds the 16T-limit of our measurements but $J_c$(16T) clearly increases by more than one order of magnitude in the CaF-samples. Moreover, the multilayer deposition appears more effective in increasing $J_c$ in the $CaF_2$ case than in LSAT case. In fact, in the films on LSAT, apart from the low field region for H//$ab$ (affected by the presence of wide plate-like precipitate in LSAT-M),[25] there is little difference in $J_c$(H) between single and multilayer films at this temperature, whereas CaF-M performs better than CaF-S in both field directions over the whole field range. Since these high temperature data are particularly sensitive to differences of $T_c$, CaF and LSAT data are also compared at the same reduced temperature (T/$T_c$~0.62) in Figure 3 (right panels). At reduced temperature, the differences in $J_c$(H) between the two single layer films are smaller: although the CaF-S still has more than 70% higher $J_c$ at self-field, LSAT-S actually performs similarly to (or slightly better than) CaF-S in the intermediate field range above 1-1.5 T. The CaF-S advantage appears again approaching the irreversibility field with $\mu_0 H_{Irr}$ about 2T larger than in LSAT-S. In contrast, the multilayers definitely have different properties: in LSAT-M, the increased pinning effectiveness is relevant only in a narrow field range for H//$ab$ whereas CaF-M has larger $J_c$ over the whole field range in both field orientations.

Low temperature characterization at 4.2 K was performed up to 35 T (CaF-samples) and 45 T (LSAT-samples) (Figure 4). The in-field behaviour of the LSAT and CaF-films appears quite different. In LSAT-M $J_c$(H//$ab$) is improved with respect to LSAT-S up to 30-35T and then the curves merge, whereas $J_c$(H//$c$) of LSAT-M is larger over the whole field range with an increase of $\mu_0 H_{Irr}$ of more than 6 T to 41 T. In the CaF-M case, $J_c$ is significantly increased with respect to CaF-S in both directions, and $J_c$(20T) exceeds 0.27 and 0.42 MA/cm$^2$ for H//$c$ and H//$ab$, respectively. The pinning force density more clearly reveals the different effectiveness of the multilayer depositions. Figure 5 shows that the maximum of $F_p$ in LSAT-M is increased by ~20% for H//$c$ and by ~30% for H//$ab$ respect to LSAT-S with a widening of the LSAT-M maximum due to overlapping pinning mechanisms [25]. In contrast, the increase of $F_p$ in CaF-M compared to CaF-S is more than 60% in both directions and the maxima exceed 84 GN/m$^3$ at 22.5 T for H//$ab$ and 70 GN/m$^3$ at 10 T for H//$c$. The angular dependence of $J_c$ (Figure 6) reveals that CaF-S has a significant $c$-axis peak, despite the absence of nanorods and also that it shows a relatively sharp $ab$-peak, whereas CaF-M has a strong $J_c$ increase in every field orientation with a decrease of the $J_c$ anisotropy compared to CaF-S. For instance at 5T the enhancements along the $c$-axis and $ab$-plane orientations are 77 and 96%, respectively, whereas the $J_c$ minimum increases by 146% reducing the $J_c$-anisotropy (=$J_{c,Max}/J_{c,Min}$) from 1.8 in CaF-S to 1.3 in CaF-M.

## Discussion

The microstructures of the films deposited on CaF$_2$ are unexpectedly different from those deposited on LSAT[25] and they also differ from other very clean Co-Ba122 films grown in ultra-high-vacuum conditions on CaF$_2$.[37] LSAT-S has, in fact, a significant amount of defects containing oxygen due to both the growth conditions and the presence of residual oxygen in the Co-Ba122 target. Despite being similarly grown, CaF-S has no oxygen-containing defects. The multilayer CaF-M was grown alternating the same Co-Ba122 target and a BaO-enriched Co-Ba122 target with the explicit intention to generate oxygen-rich defects: however this deposition produces a Ba-122 layer with fewer defects that LSAT-S and a clear defects gradient through the thickness. The complete absence of secondary phase nanorods or precipitates in CaF-S, the low density of defects in the CaF-M with a clear density gradient through the Ba122 layer, and the presence of oxygen in both the reaction layer and the top part of the substrate, all indicates that the reaction layer and substrate absorbed the oxygen, preventing the formation of all (CaF-S) or part (CaF-M) of the Ba-O precipitates. Moreover, there is a clear difference in the thickness of the transition layers in the single and multilayer films, probably due to two reasons. The deposition time is definitely longer for the CaF-M because of the time needed to switch between the two targets which means the sample remains at high temperature longer giving more time for the reaction to occur. Second, the target used for the insertion layers has extra BaO: this additional oxygen likely generates a stronger reaction and thicker transition layer. Interestingly the presence of this transition layer neither affects the Ba122 crystalline quality nor decreases the $T_c$-enhancing strain as demonstrated by the narrow FWHM and sharp $T_c$ transition. This suggests that transition layer formation occurs in a late stage of the deposition and/or during the cooling when the Ba122 phase is already stable and well crystallized.

The self-field $J_c$ of the CaF films exceeds 1MA/cm$^2$ above 16 K, which is significantly larger than previously reported[37] despite a similar $T_c$ ($T_{c,90}$ ~ 26.9-27.5 K versus 26.9-27.9 K in ref. 37). Considering the large difference in thickness between our thick films (>330nm) and Kurth *et al.*'s thin films (50-85nm), it is possible that our films are less strained ($c$=13.09 Å in CaF-S and $c$=13.17-13.19 Å in ref. 37). Perhaps the very highly strained, very thin films develop a partially compromised connectivity. In CaF-S the presence of the wide $c$-axis peak (Figure 6) is quite surprising considering that the only observed defects in the Ba-122 layer are dislocations with $B_\phi$(H//$c$) < 3 T. However the large width of the peak is probably due to the dislocation splay that can reach angles up 45° with respect to the $c$-axis.[28] The temperature-field dependence of the CaF-S $c$-axis peak can shed more light on the different pinning mechanisms involved. Figure 7 shows the angular dependences of $J_c$ normalized to the *ab*-peak at different temperatures and at 1 and 4 T (i.e. below and above the dislocation matching field). At 1 T the curves rescale over a wide angular range and the highest temperature curve has the most intense $c$-axis peak. In contrast, at 4 T the curves only rescale close to the *ab*-peak and the lowest temperature curve has the most intense *c*-axis peak. It is worth noting that this behaviour is definitely different from that observed in LSAT-S where at every field the *c*-axis peak becomes more and more intense with decreasing temperature. As described in refs. 23 and 25, the diameter of the nanorods present in LSAT-S is comparable to $2\xi_0$, allowing decreasing temperature to provide a better match between the defect size and the vortex core, thus generating a stronger pinning at lower temperature. In CaF-S the 1 and 4 T behaviours can be understood in terms of a temperature dependent change in the dominant pinning mechanism. The dislocations in CaF-S have much smaller size than the vortex core at every temperature, producing a relatively weak pinning compared to the LSAT-S nanorods. However at 1 T (below $B_\phi \sim 3$ T), the vortex-vortex interaction is small and the dislocations outnumber the vortices, providing an effective, although weak, vortex pinning. The small temperature difference in the amplitude of the *c*-axis peak on going from 8 to 16 K is likely related to vortex rigidity changes: at high T the vortices are less rigid and they can be more effectively pinned. On the other hand at 4 T (above $B_\phi$), vortex-vortex interactions increase the vortex stiffness and the dislocations alone cannot efficiently pin the vortices at high temperature, producing a less pronounced *c*-axis peak. Decreasing temperature however generates a more intense and wider peak, suggesting that an additional isotropic pinning mechanism is playing a role. Another hint of an additional contribution from a more isotropic pinning mechanism at low temperature and high field is given by the width of the *c*-axis peaks at 8 K: in fact, despite the intensity of 8 K *c*-axis peaks at 1 and 4 T are substantially unchanged (Figure 7), the peak at 4 T is ~10° wider than at 1 T and it extends all the way from 90° to 170° (only 10° from the *ab* plane). Considering the clean microstructure of this sample (Figure 1(a)), the new mechanism in play could be due to (unobserved) point defects that act as weak pinning centres at low temperature. Despite the superposition of different pinning mechanisms, this hypothesis is confirmed by the temperature dependence of the normalized pinning force curves shown in Figure 8. In fact the peak position moves from ~$0.2H_{Irr}$ at high temperature toward ~$0.3H_{Irr}$ at low temperature, which is close to the typical value for pinning by point defects ($0.33H_{Irr}$). The nature of the point defects is actually unknown and their presence is hard to identify since in all other samples there is a much stronger contribution from strong pinning: point defects might be defects/vacancies present in all samples (both on LSAT and CaF) but producing a much smaller contribution with respect to nanorods, nanoparticles and large precipitates, or they might be present only in the CaF samples because of the large strain induced by the substrate.

In Figure 5 we observed that the multilayer deposition on $CaF_2$ appears to be significantly more effective in increasing the pinning performance than on LSAT. We might think that such an increase is partially related to the fact that a single-layer on $CaF_2$ is much cleaner than the one on LSAT, so they actually start from a different level of pinning in the single layers. However the reduced temperature curves shown in Figure 9 reveals a more complicated scenario. The $J_c(T/T_c)$ values for LSAT-S and CaF-S (open symbols) are relatively similar suggesting that, despite their different pinning landscapes, the main difference in the single layers is their $T_c$ difference. However, in the multilayer samples (closed symbols in Fig.9), $J_c(T/T_c)$ shows a huge difference between the LSAT and $CaF_2$ cases: the $J_c$ increase going from CaF-S to CaF-M is 2 or 3 times larger than from LSAT-S to LSAT-M. This result is very surprising considering the much lower density of artificially introduced defects in CaF-M compared to LSAT-M. The 2-3 times more effective pinning could be the results of a combined effect of the artificial defects, visible in the TEM cross-section (Figure 1b-c), and the (invisible) point defects that are very effective in CaF-S (point defects seem absent or to have a negligible effect in the LSAT films). Another possible explanation would be the different compositions of the insertion layers in CaF-M and LSAT-M: the undoped Ba122 target used as insertion layer for LSAT-M could in fact induce a Co-deficiency in the superconducting matrix surrounding the defects, locally reducing $T_c$ and $J_c$ or even increasing the effective defect size by reduction of the effective film cross-section. This issue is not present in CaF-M since the insertion layer is Co-doped. Finally, a combination of all these effects is also possible to explain the larger effectiveness of the pinning performance in CaF-M.

A different method to introduced pinning centres was recently investigated in the P-doped Ba122 phase,[30] which has a higher bulk $T_c$ (30-31 K). $BaZrO_3$ was added in the P-doped Ba122 target inducing the formation of BaZrO$_3$ nanoparticles. However, these P-doped films were not strained and their $T_c$ was lower than the bulk $T_c$ and actually similar to our CaF samples. Despite the BaZrO$_3$ precipitates being effective pinning centres with a clear $J_c$ enhancement with respect to the clean P-Ba122, the maximum $F_p$ of 59 GN/m$^3$, though high, was still less than in the best of these films. In Figure 5 the $F_p$ curve of a high-$J_c$ Nb$_3$Sn wire[40],[41] is shown as a comparison together with the our results on Ba122. The very high $F_p$ and the fact that $F_{p,Max}$ occurs at 22.5 T, not at 5-6 T as in Nb$_3$Sn, explains the reason Ba122 is of potential interest for high-field applications. Despite the small anisotropy of Ba122, which is absent in Nb$_3$Sn, the in-field performance of Ba122 widely exceeds that of Nb$_3$Sn above 10T, thanks to an irreversibility field almost twice as large and to the strong pinning properties that shift the maximum of $F_p$ to high field.

To conclude, in this paper we investigated the superconducting properties of single- and multilayer Co-Ba122 thin films grown on CaF$_2$ and we compared the results to films on LSAT. We confirmed the significant improvement (~3.2-3.3K) of the critical temperature in the film on CaF$_2$ already observed in ref. 37 and we demonstrated that the multilayer deposition has no detrimental effect on $T_c$.

We found that deposition on CaF$_2$ produces much cleaner films compared to films on LSAT, since on CaF$_2$ the reaction layer and the substrate absorb oxygen preventing the formation of the high density of defects present in the LSAT case. The result of this oxygen reaction is that in the single layer on CaF$_2$ (CaF-S) the only visible defects are dislocations. Moreover, the different trend of $J_c(\theta)$ below and above $B_\phi$ in CaF-S at intermediate temperature suggests an increasing contribution of point defects with decreasing temperature. The multilayer film on CaF$_2$ (CaF-M), grown with insertion layers enriched in BaO, does have defects including short nanorods, round particles and flat particles similar in size but with a lower density to those in films on LSAT. Despite the different pinning mechanisms that affect the shape of the angular dependence of $J_c$, the single layers on both substrates roughly have similar $J_c$ at the same reduced temperature (Figure 9). In contrast the multilayer deposition on CaF$_2$ is 2-3 times more effective in increasing $J_c$ with respect to the corresponding single layer than in the LSAT case. This results in a record high $F_p$ of 84 and 70 GN/m$^3$ in Ba122 phases (H//*ab* and H//*c*, respectively), significantly larger than reported previous values of 53 GN/m$^3$ in the Co-Ba122 phase[25] and of 59 GN/m$^3$ in the higher-$T_c$ P-Ba122 phase.[30] Applications interest is strengthened by their notably better in-field performance than in Nb$_3$Sn.

**Methods**

**Target preparation, film growth and TEM characterization.** The four Ba122 films studied in this paper were grown by PLD as described in Refs. 18 and 19 using a Co-Ba122 target synthesized using Ba$_3$As$_2$ as the barium source in order to minimize the oxidation and better control the oxygen content:[25] the elements were balanced to Ba$_{0.95}$Fe$_{1.84}$Co$_{0.16}$As$_{2.1}$ (preparation details are in ref. [42]). The other two targets were synthesized to be used in the multilayer deposition to introduce defects in the films. The first of these was an undoped Ba-122 (un-Ba122) target used for the LSAT-M film and it was prepared employing pure elements as reactants: because of the presence of unreacted Ba after the target heat treatment, oxidation occurred more likely than by using Ba$_3$As$_2$, producing oxygen-rich defects in the final films (see ref.[25] for more details). The second of these was used for the insertion layer in the CaF-M film and was prepared similarly to the Co-Ba122 target but with the addition of BaO (BaO-doped Co-Ba122 target). All targets were heat treated at 1120°C for 12 hours under a hot isostatic press at 193 MPa.

Two single layer films were grown by pulsed laser deposition (PLD) with the Co-Ba122 target: one on (001)-oriented (La,Sr)(Al,Ta)O$_3$ substrates with an intermediate template of 100 unit cells (u.c.) of SrTiO$_3$ (named LSAT-S)[18,25,39] and one on (001)-oriented CaF$_2$ (named CaF-S). The multilayer film on STO-templated LSAT consisted of by 24 Co-Ba122/un-Ba122 bilayers (thickness ratio t$_{Co-Ba122}$:t$_{un-Ba122}$ = 4:1) reaching a total thickness of ~400 nm (film named LSAT-M).[25,39] The multilayer film on CaF$_2$ was deposited with 24 bilayers of Co-Ba122/BaO-doped Co-Ba122 (thickness ratio t$_{Co-Ba122}$:t$_{BaO-doped\ Co-Ba122}$ = 4:1) reaching a total thickness of ~350 nm.

Microstructural characterization was performed in a JEOL ARM200cF transmission electron microscope (TEM) equipped with an electron energy loss spectroscopy (EELS).

**Transport properties.** Transport characterizations ware performed on 40 µm wide and 1mm long bridges fabricated by laser cutting. The I-V curves were measured in the maximum Lorentz force configuration in a physical property measurement system (PPMS) up to 16 T and at the DC facility at the National High Magnetic Field Laboratory (NHMFL) in Tallahassee up to 45 T. The critical current was determined by a 1µV/cm criterion.

**Acknowledgements**

This work was supported at FSU by NSF DMR-1306785, the State of Florida and by the National High Magnetic Field Laboratory which is supported by the National Science Foundation under NSF DMR-1157490, and the State of Florida. Work at the University of Wisconsin was supported by the DOE Office of Basic Energy Sciences under award number DE-FG02-06ER46327.

**Author Contributions**

C.T. carried out the low and high field transport measurements and prepared the manuscript. F.K. performed the HRTEM and EELS characterizations. S.L. deposited the Ba-122 thin films. J.Jiang and J.D.W. synthetized and characterized the targets. J.Jaroszynski performed the high field transport measurements. E.E.H., C.B.E and D.C.L. directed the research and contributed to manuscript preparation. All authors discussed the results and implications and commented on the manuscript.

**TABLE I.** Thin film structures and superconducting properties

| Sample name | Substrate | Description and thickness | $T_{c,0}$ (K) |
|---|---|---|---|
| CaF-S | $CaF_2$ | Single-layer, Co-Ba122 (330nm) | 25.4 |
| CaF-M | $CaF_2$ | Multi-layer (350nm) Co-Ba122 / BaO-doped Co-Ba122 | 26.0 |
| LSAT-S | $(La,Sr)(Al,Ta)O_3$ | Co-Ba122 (400nm) | 22.1 |
| LSAT-M | $(La,Sr)(Al,Ta)O_3$ | Multi-layer (400nm) Co-Ba122 / un-Ba122 | 22.8 |

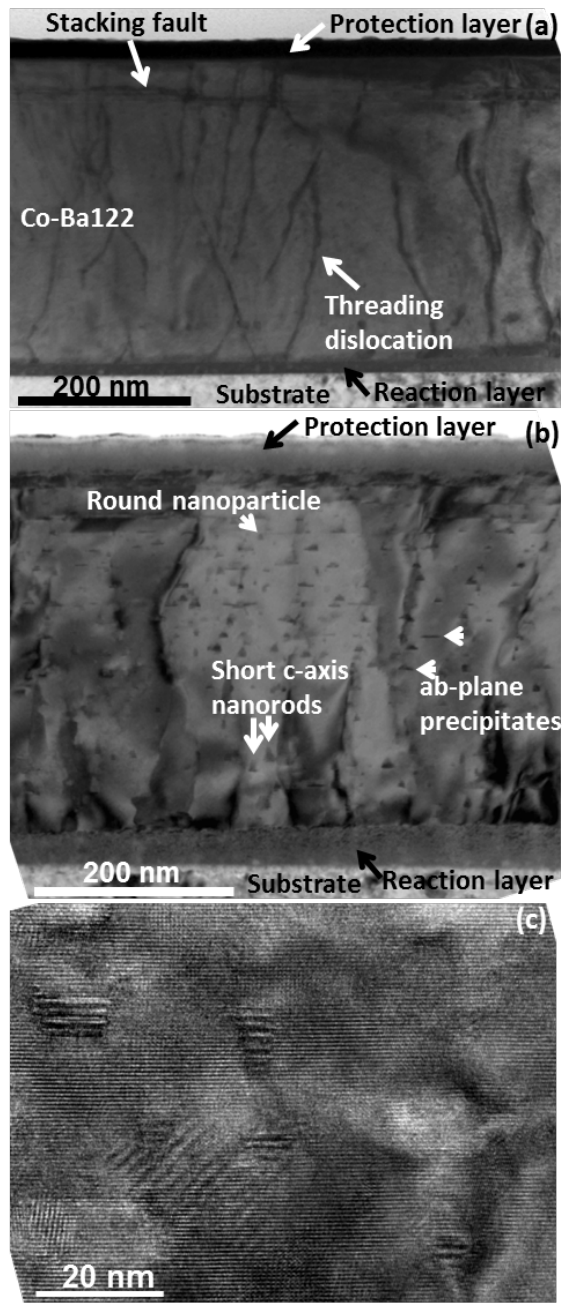

**Figure 1 TEM cross sections of Co-Ba122 thin films on CaF$_2$.** (a) Single-layer Co-Ba122 film with a very clean microstructure with only threading dislocations and few stacking faults. (b) Multilayer Co-Ba122 films with short *c*-axis nanorods, round nanoparticles and *ab*-arranged precipitates: this sample present a defects gradient across the thickness. (c) High magnification of the few defects present in the multilayer film shown in (b).

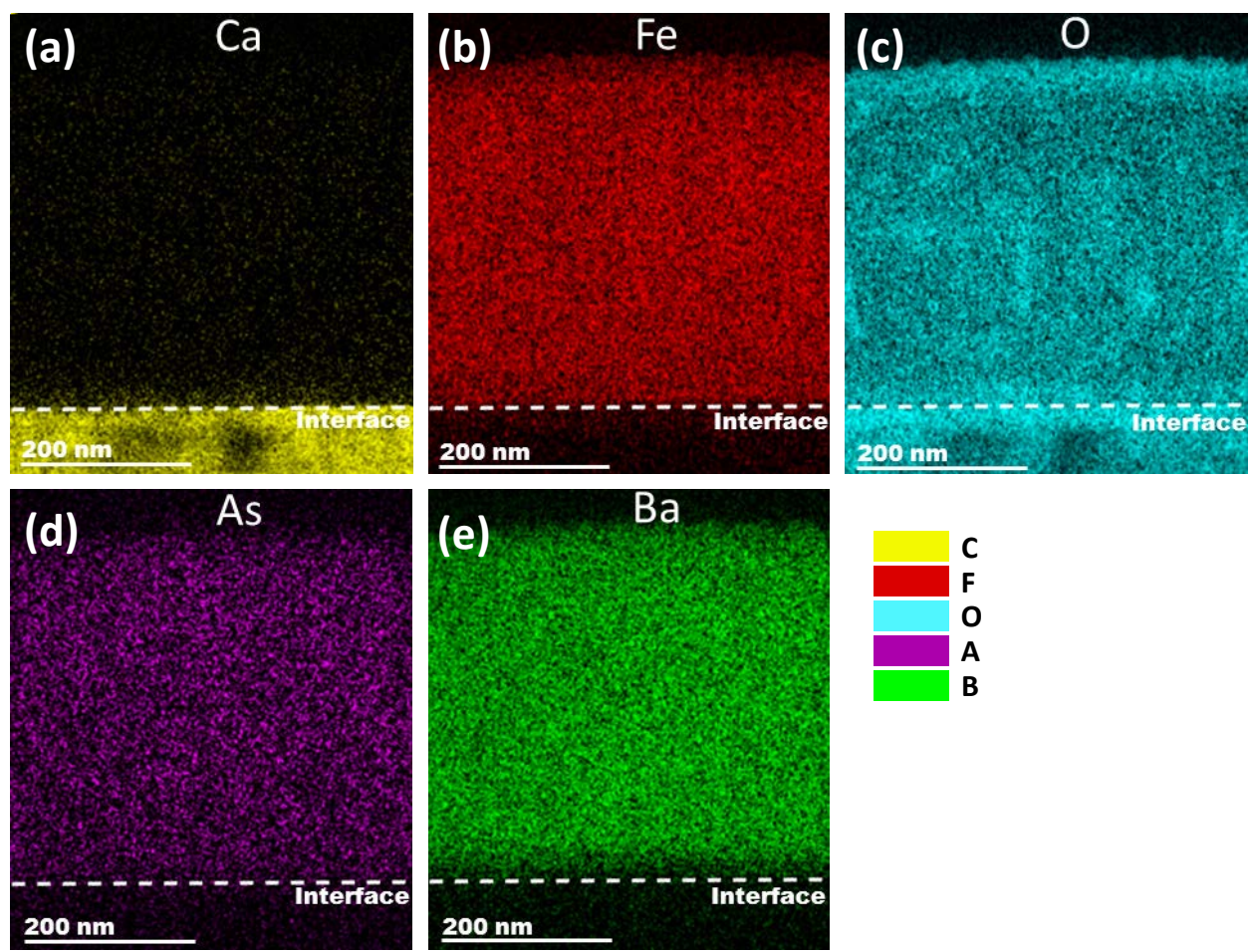

**Figure 2 Elemental maps of the multilayer Co-Ba122 thin films on CaF$_2$.** The Fe and As map are uniform across the Ba122 and reaction layers. Ba and Ca maps indicate an interdiffusion of the two element in the reaction layer. O map reveals an high concentration of oxygen in the reaction layer and in the top part of the substrate (above and below the interface) and presence of oxygen in the Ba122 layer. The interface is the top of the CaF$_2$ substrate.

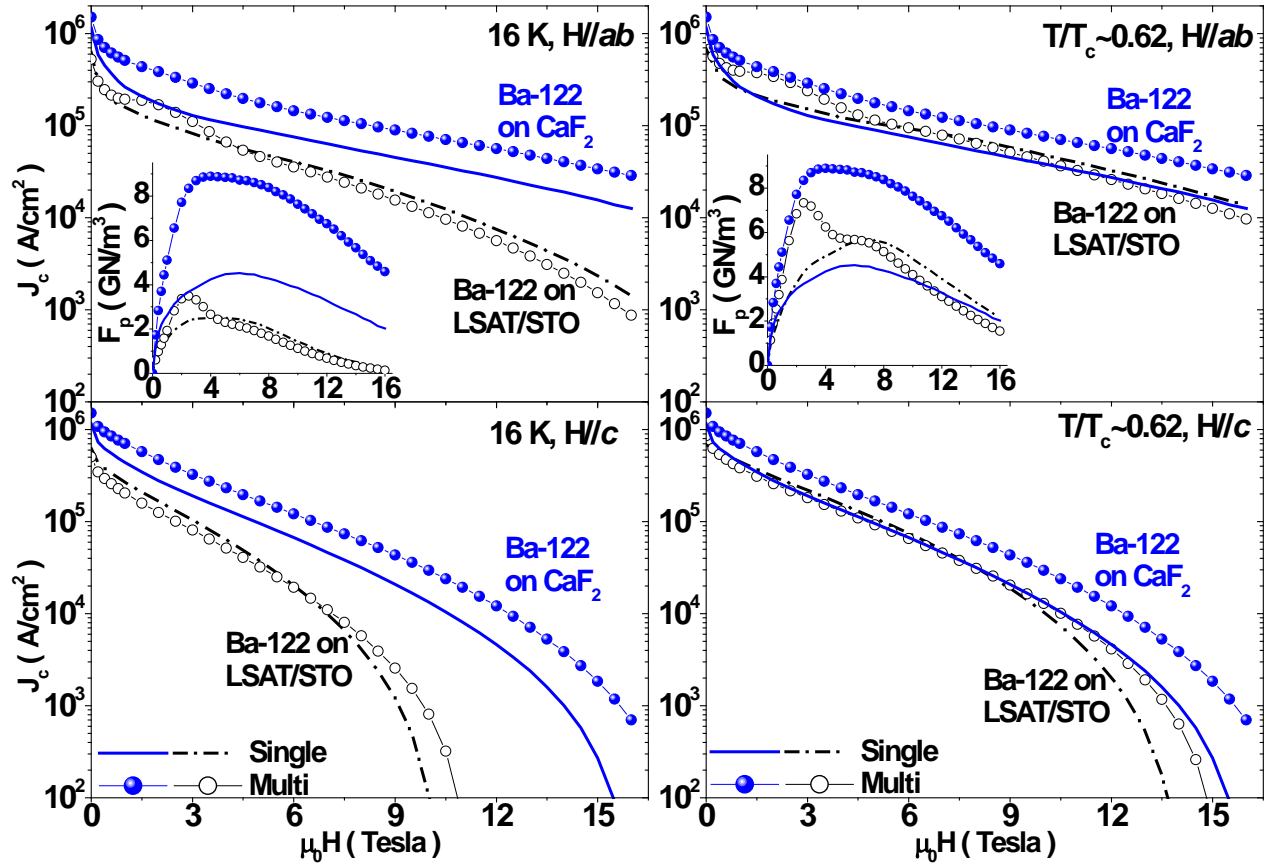

**Figure 3 $J_c$ as a function of field for single and multilayer Co-Ba122 films at high temperature.** $J_c$ data at 16 K (left) and at T/$T_c$~0.62 (right) along the two main field orientations for the films deposited on CaF$_2$ (blue solid lines and symbols) and LSAT/STO (black dashed lines and open symbols). In the inset the $F_p$(H) curves for H//*ab*.

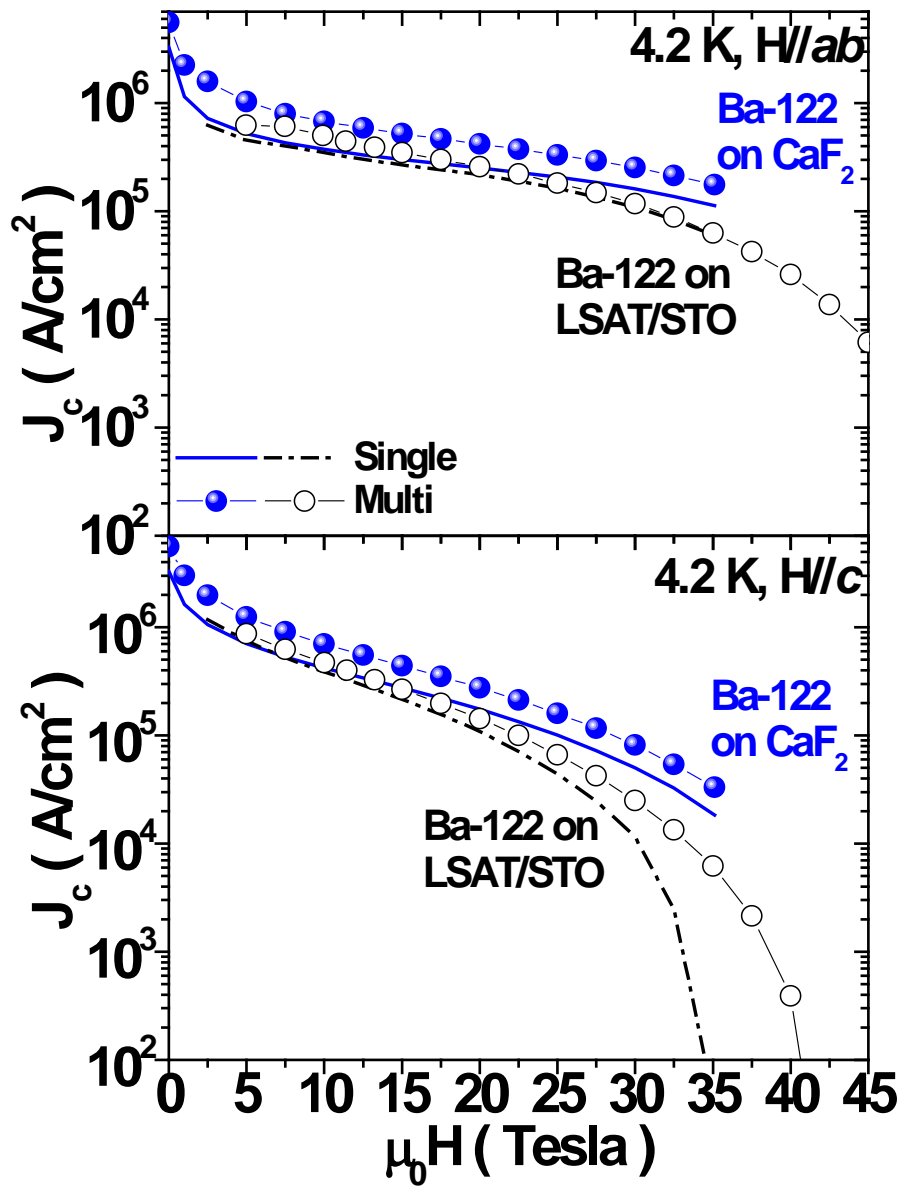

**Figure 4 High-field $J_c$ at 4.2 K for single and multilayer Co-Ba122 films.** $J_c$ data along the two main crystallographic orientations the films deposited on $CaF_2$ (blue solid lines and symbols) and LSAT/STO (black dashed lines and open symbols).

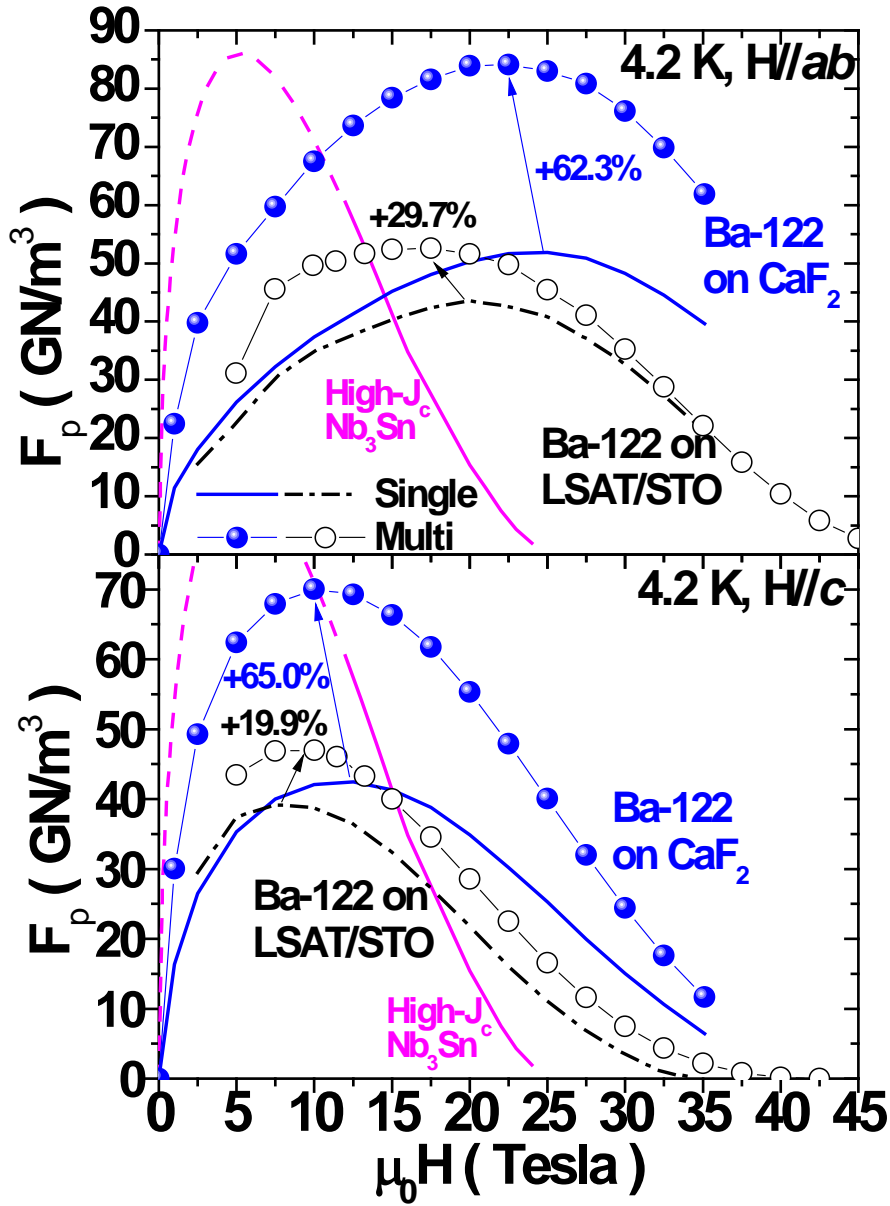

**Figure 5 High-field $F_p$ at 4.2 K for single and multilayer Co-Ba122 films.** $F_p$ data along the two main field orientations for the films deposited on on CaF$_2$ (blue solid lines and symbols) and LSAT/STO (black dashed lines and open symbols). The magenta line represents $F_p$ at 4.2 K of a high-$J_c$ Nb$_3$Sn wire [40,41].

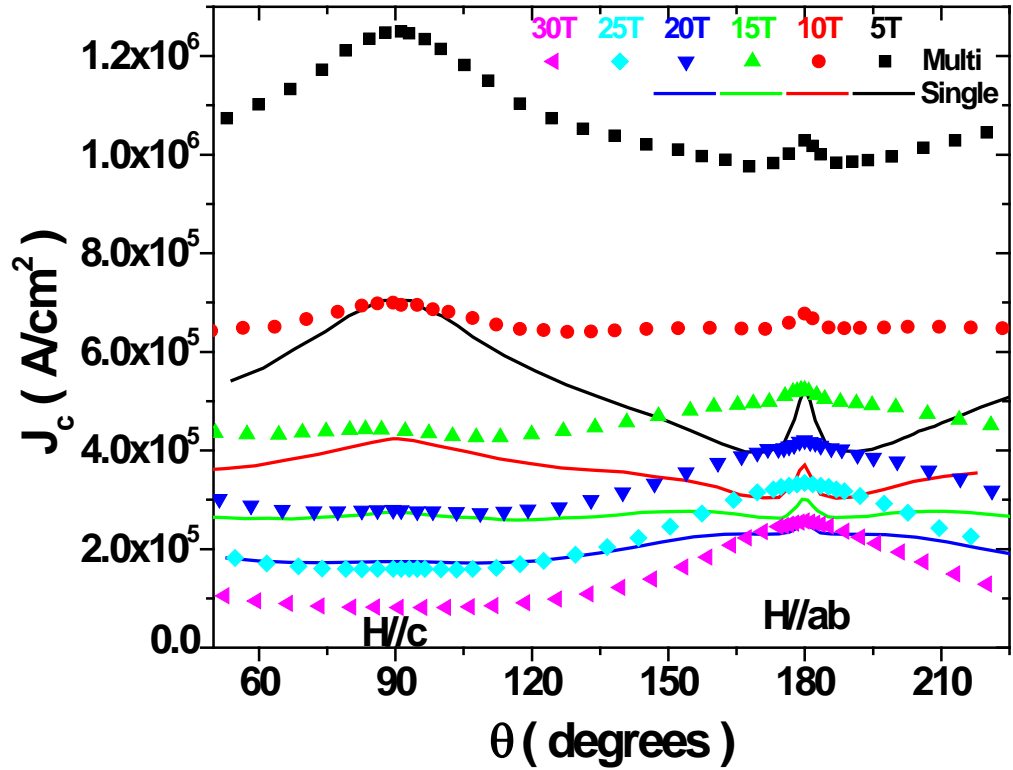

**Figure 6 High-field angular dependences of $J_c$ at 4.2 K for the single and multilayer films on CaF$_2$.** A significant increase of $J_c$ is found in the multilayer sample compared to the single-layer. The multilayer deposition also strongly decreases the $J_c$ anisotropy ($J_{c,\text{Max}}/J_{c,\text{Min}}$) (e.g. from 1.8 in the single-layer to 1.3 in the multilayer at 5T).

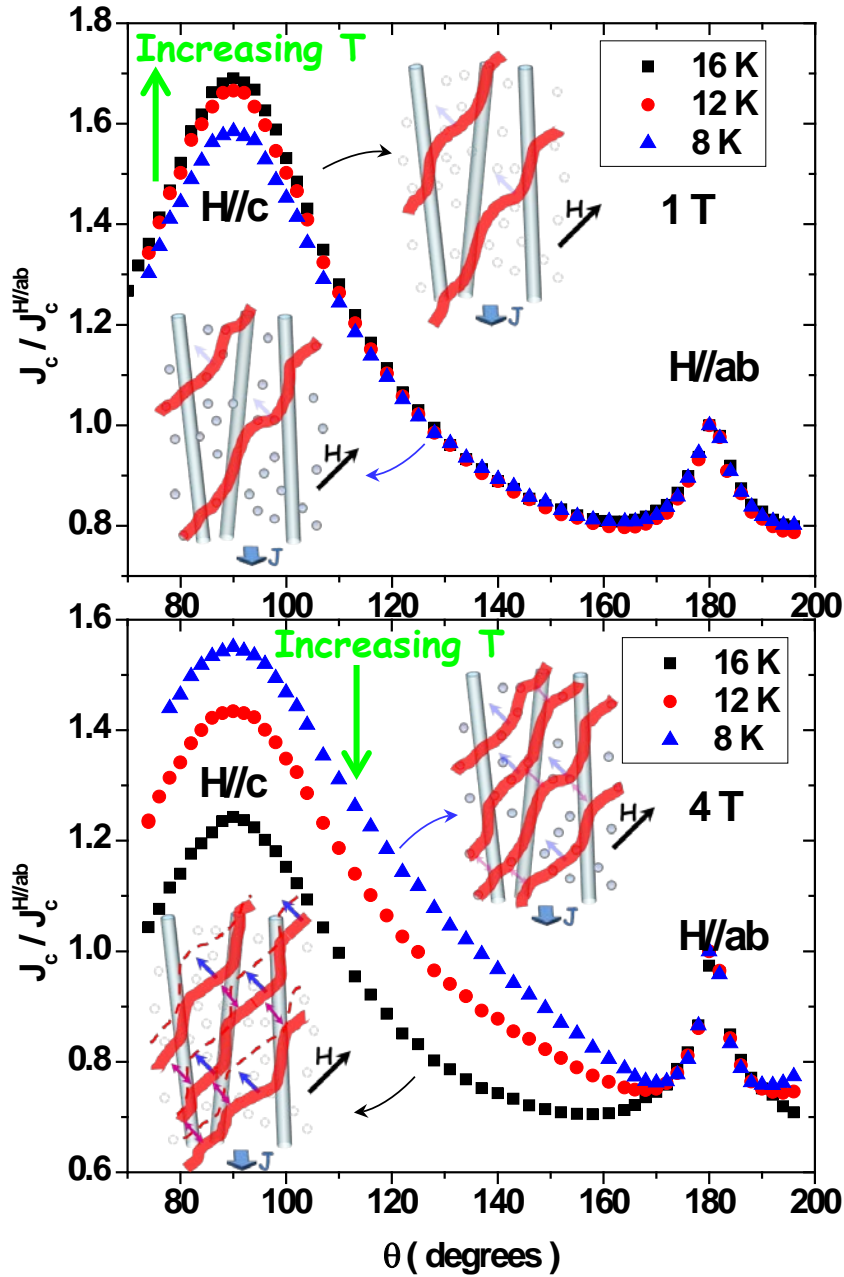

**Figure 7 Angular dependences of $J_c$ normalized to the *ab*-peak for the single-layer Co-Ba122 film on CaF$_2$.** The normalized $J_c$ data are reported for the 8-16 K range at 1 and 4T (below and above the dislocation matching field), showing the change in pinning effectiveness with temperature. In the inset, sketches of the different pinning landscape with vortex lines (red lines), dislocations (gray cylinders), effective and ineffective point defects (solid and dashed dots); the blue and purple arrows represent the Lorentz force and the vortex-vortex interaction.

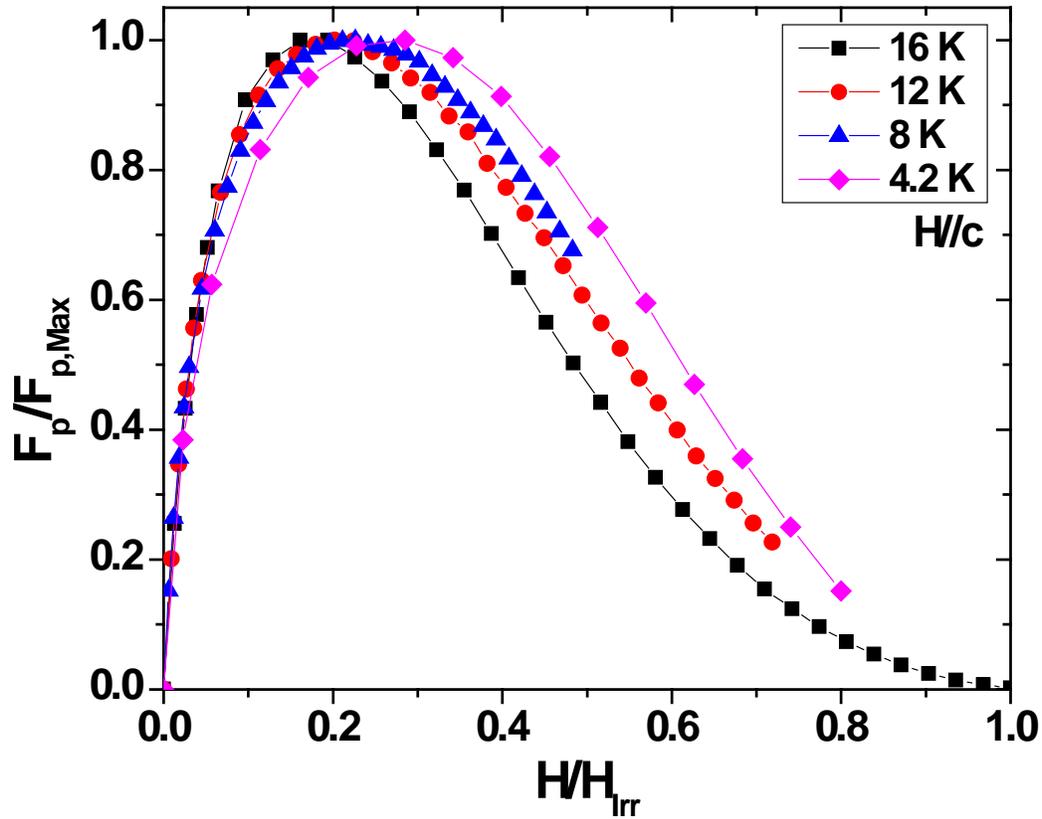

**Figure 8 Normalized pinning force curves as a function of reduced field for the single-layer Co-Ba122 film on CaF$_2$.** The normalized $F_p$ data are reported for the 4.2-16 K range showing the change of the dominant pinning mechanism with temperature.

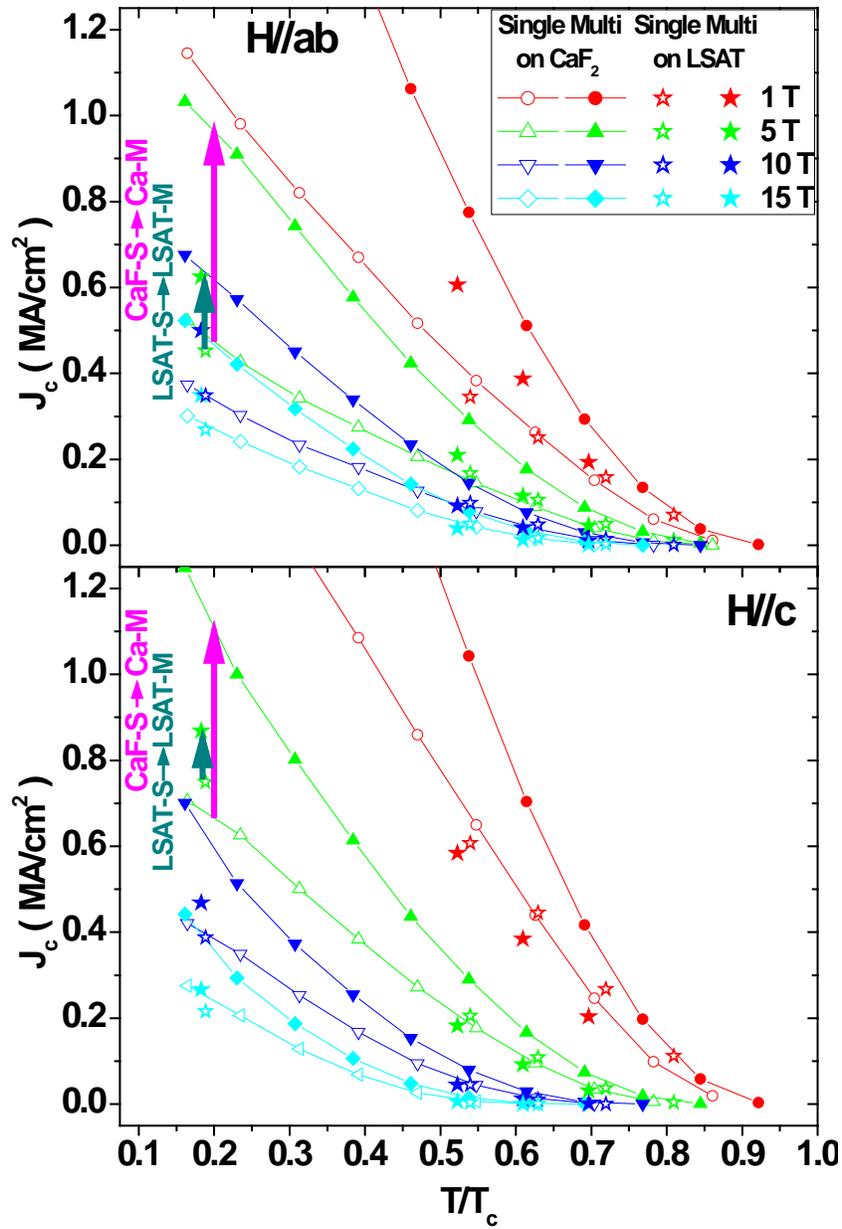

**Figure 9 Temperature dependences of $J_c$ for single and multilayer Co-Ba122 films.** The multilayer deposition on $CaF_2$ (lines and closed symbols) increases $J_c$ with respect to the single-layer on $CaF_2$ (lines and open symbols) 2-3 times more effectively than in the LSAT case (closed and open stars for the multi- and single layers, respectively). For instance, the arrows show the $J_c$(5T) increase going from LSAT-S to LSAT-M (dark green) and from CaF-S to CaF-M (magenta) in the low temperature region.